\documentstyle[preprint,prb,aps]{revtex}
\begin{document}
\draft
\title{Phase Transitions and Vortex Line Entanglement in a Model
High Temperature Superconductor}
\author{Ying-Hong Li\cite{yl}}
\address{Institute for Theoretical Physics,
       University of Utrecht,
       P.O. Box 80.006, 3508 TA  Utrecht,
       The Netherlands}
\author{S. Teitel}
\address{Department of Physics and Astronomy, University of Rochester,
 Rochester, New York 14627}
\date{\today}
\maketitle
\begin{abstract}
% insert abstract here
We carry out Monte Carlo simulations of the uniformly frustrated three
dimensional XY model, as a model
for vortex line fluctuations in high temperature superconductors in an
applied magnetic field.
We show, comparing systems of different size, that
there are two distinct phase transitions.  At a lower $T_{c\perp}$,
the vortex lattice melts, and coherence is lost in planes perpendicular
to the magnetic field.  At a higher $T_{cz}$, a vortex tangle
percolates throughout the system, and coherence is lost parallel to the
magnetic field.  Cooling below $T_{cz}$, high energy barriers for vortex
line cutting lead to an entangled glassy state.
\end{abstract}

\narrowtext
\section{Introduction}

In several earlier works\cite{r1,r2} we have introduced the three dimensional
uniformly frustrated XY model\cite{r3}, as a phenomenological
model for studying phase transitions,
and the effects of vortex line fluctuations, in
the mixed state of high temperature
superconductors in a uniform applied magnetic field.
This model applies in the strongly type-II limit where the
magnetic penetration length is much greater than the average
vortex line separation, $\lambda\gg a_v$, and the magnetic
induction inside the superconductor is approximately uniform.
In our most recent work\cite{r2}, henceforth referred to as (I),
we found evidence that the system undergoes two
distinct phase transitions upon heating.  First, at the lower $T_{c\perp}$,
the vortex line lattice melts, destroying superconducting phase
coherence in directions perpendicular to the applied magnetic field;
coherence parallel to the field however remains.  Then, at the
higher $T_{cz}$, coherence parallel to the field is lost as well.

In the present work we extend the results of (I) in several major
directions: ($i$) By presenting detailed
studies of the system behavior as the system size is varied in the
directions parallel and perpendicular to the magnetic field, we show
clearly that the two distinct transitions found in (I) are
not artifacts of finite size effects.
($ii$) We show
that the upper transition $T_{cz}$ can be viewed as
a percolation-like transition, where the
vortex lines become so completely interconnected through mutual intersections,
that one may trace out a connected path of vortex line segments which
travels completely around the system in the direction $perpendicular$ to
the applied magnetic field.
($iii$) We find, in contrast to our earlier results,
that below $T_{cz}$ the energy barrier for vortex line cutting
grows so large, that cuttings are frozen out on the time scale of our
simulation, and the system can cool into an entangled glassy state as
in the ``polymer glass" picture of Obukhov and Rubinstein\cite{r4}.

The remainder of our paper is organized as follows.  In Section II we
outline our model and Monte Carlo method.  In Section III we present
the results of our simulations.  Section IIIA gives results for
the helicity modulus, which measures superconducting phase coherence.
Section IIIB gives results for the average length of vortex lines due
to thermal fluctuations.  Section IIIC analyses the entanglement of the
vortex lines by considering the winding of the field induced vortex lines
about the direction of the magnetic field.  Finite size dependencies
are investigated.  Section IIID gives results concerning the distribution
of thermally excited closed vortex rings.  Section IIIE discusses
the ``2d boson" analogue to vortex line fluctuations, as applied to
our model.  In Section IV we summarize our results and discuss the
possible connection to recent experiments.

\section{Model}

The model that we study is given by the Hamiltonian\cite{r2}
\begin{equation}
{\cal H}[\theta_i]=J_0\sum_{\langle ij\rangle}V(\theta_i-\theta_j-A_{ij})
\label{eq:h}
\end{equation}
where $\theta_i$ is the phase of the superconducting wavefunction at site
$i$ of a three dimensional
cubic numerical mesh, the sum is over all nearest neighbor
bonds of this mesh,
\begin{equation}
A_{ij}={2\pi\over\Phi_0}\int_i^j{\bf A}\cdot d{\bf l}
\label{eq:a}
\end{equation}
is proportional to the integral of the fixed magnetic vector potential
${\bf A}$ across bond $\langle ij\rangle$ ($\Phi_0=hc/2e$ is the
flux quantum),
\begin{equation}
V(\alpha)\equiv -(T/J_0)\ln\left\{\sum_{m=-\infty}^\infty\exp\left[
-{1\over 2}J_0(\alpha -2\pi m)^2/T\right]\right\}
\label{eq:v}
\end{equation}
is the Villain interaction between neighboring sites, and the coupling is
\begin{equation}
J_0={\Phi_0^2\xi_0\over 16\pi^3\lambda^2}
\label{eq:j}
\end{equation}
where we identify the vortex core radius $\xi_0$ with the lattice constant
$a$ of our numerical mesh.  For our numerical studies we take an isotropic
uniform constant $J_0$, although this could be varied if desired to model
the effects of anisotropy or randomness.  Periodic boundary conditions
are chosen in all directions.  Numerical meshes of various sizes $L_\perp^2
\times L_z$ are studied (the subscript ``$\perp$" will refer to the
${\bf\hat x}$ and ${\bf\hat y}$ directions, transverse to the applied
magnetic field).

The approximations which lead from the familiar Landau-Ginzburg free energy
functional to the Hamiltonian (\ref{eq:h}), and their justifications in
the $\lambda\gg a_v$ limit, have been described in detail in (I).  In
the following, we study the specific case where the uniform magnetic induction
${\bf B}={\bf\nabla}\times{\bf A}$
is $f=1/25$ flux quantum per plaquette of the mesh, oriented in the
${\bf\hat z}$ direction.  The ground state is a square periodic
lattice of straight magnetic field induced
vortex lines with spacing $a_v/a=1/\sqrt{f}=5$, as shown in
Fig.\ \ref{fig1}.  Henceforth, we cite all lengths in units of $a\simeq\xi_0$,
and energies in units of $J_0$.
The total number of field induced lines is $N_v=fL_\perp^2$.

Our Monte Carlo simulations are carried out using the standard Metropolis
algorithm.  Performing the simulation in terms of the phase variables
$\theta_i$, we locate the vortex lines in any particular configuration
by computing the net phase change around every plaquette of the mesh.
We define as an intersection, or cutting, between two vortex lines
whenever we find a unit cell of the mesh which has more than one
vortex line entering and leaving.  In such a situation, we randomly assign
which exiting segment is connected to which entering segment, for
the purpose of identifying the paths of these particular lines.
Each of our data points is typically the result of $2,000$ sweeps to
equilibrate,
followed by $15,000$ sweeps to compute averages, where each sweep
refers to one updating pass through the entire numerical mesh.
Our calculations were carried out on a Sparc 10 workstation; for our
largest system, $25^2\times 200$, each temperature point took approximately
two CPU days of computation.

\section{Results}
\subsection{Helicity Modulus}

To investigate phase coherence, we compute the helicity moduli, which
give the stiffness of the system to twisted phase boundary conditions (see
(I) for derivation).  In Fig.\ \ref{fig2} we show our
results for the the helicity modulus $\Upsilon_\perp(T)$ perpendicular
to the magnetic field, and $\Upsilon_z(T)$ parallel to the magnetic
field, for lattices of fixed size $L_\perp=25$, but varying $L_z=50$, $100$,
and $200$.  In Fig.\ \ref{fig3} we show $\Upsilon_\perp(T)$ and
$\Upsilon_z(T)$ for approximately equal $L_z=24$, $25$, but varying
$L_\perp=25$ and $50$.  We see clearly two transitions, with $\Upsilon_\perp$
vanishing at $T_{c\perp}\simeq 1.35$, and $\Upsilon_{z}$ vanishing at
$T_{cz}\simeq 2.6$.  Comparing the results from different $L_\perp$ and $L_z$,
finite size effects are generally seen to be small, hence
we have clear evidence for three distinct thermodynamic states.  The middle
state is one in which superconducting phase coherence is destroyed in
planes perpendicular to the magnetic field, but coherence is preserved
in the direction parallel to the magnetic field.  Comparing the results
for heating versus cooling, we see only a small hysteresis in $\Upsilon_z$,
however hysteresis in $\Upsilon_\perp$ increases with increasing $L_z$.
We will see that this hysteresis in $\Upsilon_\perp$ is related to
the entanglement of the vortex lines as they cool into a glassy state.

\subsection{Vortex Line Lengths}

As a first measure of the amount and nature of vortex line fluctuations,
we consider the average density of vortex line segments in different
directions.  In the ground state, the total length of vortex lines is
$N_vL_z$ along ${\bf \hat z}$, while zero along ${\bf \hat x}$
and ${\bf \hat y}$.
In Fig.\ \ref{fig4} we plot the vortex line length density
$\Delta l_\mu$ versus temperature,
where $\Delta l_\mu$ is defined as the total length of vortex line segments
due to fluctuations  (ie. in excess over the ground state value)
in direction ${\bf\hat \mu}$, normalized by
$N_vL_z$.  $\Delta l_\mu =1$ represents an
excess vortex line length equal to that of the straight field induced
lines at $T=0$.  We see that $\Delta l_{x,y}\gg\Delta l_z$ in the vortex
line lattice phase below $T_{c\perp}$, as well as for much of the
vortex line liquid phase between $T_{c\perp}$ and $T_{cz}$.  This
indicates that in these regions, the dominant fluctuations are
directed transverse fluctuations of the magnetic field induced vortex
lines, as shown schematically in Fig.\ \ref{fig5}$a$. Near and above $T_{cz}$
however, we find that $\Delta l_{x,y}\simeq \Delta l_z$.  We will see that
this is due to the proliferation of closed vortex ring excitations
as shown in Fig.\ \ref{fig5}$c$.  Above $T_{cz}$,
$\Delta l_\mu > 1$, and the total vorticity is dominated by the contribution
from fluctuations.  We will see that this region is an interconnected
tangle of vortex line segments, with no unambiguous separation
between field induced lines and thermally excited rings.

\subsection{Entanglement}

To consider the entanglement of the field induced vortex lines, we
make use of the periodic boundary condition which is imposed
along the direction of the magnetic field ${\bf\hat z}$.  If
$\{ {\bf r}_{\perp i}(z)\}$ are the positions
in the $xy$ plane where the field induced vortex lines intersect the
plane at constant $z$, then the set of points $\{ {\bf r}_{\perp i}(0)\}$
must be identical to the set of points $\{ {\bf r}_{\perp i}(L_z)\}$.
If we view this periodic boundary condition along ${\bf\hat z}$
as representing the circumference of a three dimensional torus, then
the magnetic field induced vortex lines will divide into distinct
connected groups, each of which makes a certain number of windings
around the system in the ${\bf\hat z}$ direction before closing
back on itself.  A group making a winding $m$ would consist
of the $m$ lines $i_1, i_2, ..., i_m$ satisfying the condition,
${\bf r}_{\perp i_1}(0) = {\bf r}_{\perp i_2}(L_z)$,
${\bf r}_{\perp i_2}(0) = {\bf r}_{\perp i_3}(L_z)$, $...$,
${\bf r}_{\perp i_m}(0) = {\bf r}_{\perp i_1}(L_z)$.
For example, in Fig.\ \ref{fig6}, we show a configuration
with two lines of winding $m=1$, two groups of
lines with winding $m=2$, and one group of lines with winding $m=3$.
Qualitatively, when a configuration contains
only windings with $m=1$, we say that it is ``unentangled".  When a
configuration contains many windings with large values of $m$, it is
highly entangled.
To characterize the degree of entanglement, we compute
the average distribution $n(m)$ of the total number of lines $n$ which
participate in windings of value $m$.  $\sum_m n(m)=N_v$,
the total number of field induced vortex lines.

In Fig.\ \ref{fig7} we plot versus $T$ for several different system
sizes, the ratio $R\equiv n(1)/N_v$ of lines which make a winding of
$m=1$.  $R=1$ indicates a completely unentangled set of lines.  We see
that for all sizes upon $heating$ from the ground state, $R\simeq 1$
stays constant until about $T\simeq 2.0>T_{c\perp}$, then decreases
to its high $T$ limit at $T_{cz}$.
Upon $cooling$ however, $R$ starts to rise
below $T_{cz}$ and saturates around $T_{c\perp}$ to a value $R\le 1$,
dependent on system size.  Only for our shortest system, $L_z=50$, do we find
disentanglement, ie. $R=1$ upon cooling.  For all larger $L_z$,
the lines remain trapped in a non-equilibrium entangled state upon
cooling (in (I) our system size was $L_z=24$,
hence we failed to see the entanglement below $T_{cz}$ that we now find).
The degree of this entanglement increases (ie. $R$ decreases)
with increasing $L_z$.  As this cooled state is not in equilibrium,
it is unclear if the low temperature value of $R$ may vary with independent
coolings, or if it may strongly depend on the rate of cooling.  To test
this, we have carried out four independent coolings of the system size
$L_\perp=25$, $L_z=100$, and find the $T\to 0$ values of $R=0.64$,
$0.68$, $0.75$, and $0.79$.
The second of these coolings was carried out using twice the number of
Monte Carlo sweeps per temperature as for the rest of our data.  We
similarly have carried out four independent coolings of the system
size $L_\perp=15$, $L_z=100$, finding values of $R(T\to 0)=0.44$,
$1.0$, $0.41$ and $0.38$.  The
second run indicates that large fluctuations in $R$ are possible.

The strong hysteresis we find in $R$, which measures the global topology
of the lines, should be contrasted with the absence of hysteresis in the
line length densities $\Delta l_\mu$ (see Fig.\ \ref{fig4}),
which are a local measure of line fluctuations (the slight hysteresis
in $\Delta l_{x,y}$ which appears below $T\simeq 1.0<T_{c\perp}$ is due
to the extra geometrical line length needed to make a quenched entangled
state, compared to a lattice of straight lines).  This suggests that
the hysteresis in $R$ is due to the energy barrier for the cutting of
vortex lines. As $T$ decreases below $T_{cz}$,
thermal activation over this energy barrier, which is necessary to
disentangle the lines, becomes frozen out on the times scales of our
simulation.  As a measure of this energy barrier, we compute the
average number of line cuttings $N_c$ (unit cells with more than one
line entering and leaving) present in the system.  We then define the
cutting length $\xi_c\equiv N_vL_z/N_c$ as the average distance in the
${\bf\hat z}$ direction between two successive cuttings of a single
line.  We plot $\xi_c$ versus $T$ in Fig.\ \ref{fig8}.  Above $T_{cz}$
we find $\xi_c\simeq 1-2$ indicating a heavily interconnected tangle of
lines with much cutting.  As $T$ decreases below $T_{cz}$,
$\xi_c$ increases rapidly, becoming of the order of $L_z$.  The absence
of any size dependence in $\xi_c$ comparing the system with $L_z=50$
(which disentangles upon cooling) versus $L_z=200$ (which remains entangled
upon cooling) suggests that much of the cutting which determines $\xi_c$
in the region below $T\simeq 2.0$ may be due to the intersection between
field induced lines and thermally excited closed vortex rings, rather
than between two field induced lines; cuttings between field induced
lines may only be occurring on even larger length scales.  This picture
we find of cooling into a non-equilibrium entangled state is therefore
similar to the ``polymer glass transition" originally proposed by
Obukhov and Rubinstein\cite{r4}.
Comparing the data for $L_z=50$ with $L_z=200$
in Fig.\ \ref{fig2}, it is interesting to note that with respect to phase
coherence, entanglement has a noticeable effect only on the helicity
modulus $\Upsilon_\perp$; $\Upsilon_z$ seems entirely unaffected.

As a further measure of the process of vortex line entanglement, we now
consider the complete distribution of line windings $n(m)$.
In Fig.\ \ref{fig9} we show $n(m)$, for a
fixed system size of $L_\perp=25$, $L_z=200$, for various temperatures.
The results shown were obtained upon cooling the system.
We find that for all $T\ge T_{cz}\simeq 2.6$, the distribution is $n(m)=1$
for all $m$,
ie. a field induced vortex line selected at random is equally likely
to belong to a winding of any value $m$.
This result is consistent with the assumption that each vortex line $i$
is equally likely to reconnect onto any other vortex line $j$,
upon traversing the system in the ${\bf\hat z}$
direction once, ie. ${\bf r}_{\perp i}(0)={\bf r}_{\perp j}(L_z)$ is
equally likely for any $i$ and $j$.
The most likely explanation for such behavior is that above $T_{cz}$
the lines become so completely interconnected
due to cuttings, that the global path of a given line is primarily determined
by our algorithm which makes a random choice for the continuation of
the line at each individual
cutting.  When each line has sufficient cuttings with its neighbors,
the resulting line path our algorithm traces out
is equally likely to meander anywhere throughout the system.
This conclusion is supported by Fig.\ \ref{fig8} where we see $\xi_c\simeq
1-2$ for $T\ge T_{cz}$.  We will see further evidence for this later
when we consider the distribution of thermally excited vortex rings.

As $T$ decreases below $T_{cz}$ in Fig.\ \ref{fig9},
we find a steady increase in $n(m)$ at smaller $m$, compensated by
a decrease in $n(m)$ at the largest $m$.  This is as one would expect
when the connectivity of any pair of field induced vortex lines $i$ and $j$,
ie. ${\bf r}_{\perp i}(0)={\bf r}_{\perp j}(L_z)$,
becomes dominated by the thermal transverse wandering
of the lines as they pass through the system along ${\bf\hat z}$, rather
than by line cuttings.  As $T$ decreases, the transverse wandering
decreases, and the probability for near neighbor reconnections increases
with the resulting increase in $n(m)$ for small $m$.
It is interesting to note however, that even for $T$ moderately
below $T_{cz}$, there remains a wide region of intermediate $m$,
where we continue to find $n(m)\simeq 1$.  For $T=2.6-2.3$ we
believe that, even though we are below $T_{cz}$, the energy barrier
for line cutting is still sufficiently small compared to $T$, that
our data represents true equilibrium behavior (see $\xi_c\le 10$ in
Fig.\ \ref{fig8}, and the absence of hysteresis in $R$ in Fig.\ \ref{fig7},
for these values of $T$).  As $T$ is cooled below $\sim 2.3$, the
system gets trapped in some random metastable non-equilibrium tangle.

We now consider the finite size dependence of entanglement.  In
Fig.\ \ref{fig10} we plot $n(m)$ for systems of fixed $L_\perp$, but
varying $L_z=50$, $100$, and $200$.
Our data is for the fixed temperature $T=2.4$,
below $T_{cz}$ yet still high enough that we are sampling equilibrium.
We see that as $L_z$ increases, $n(m)$ approaches the $T>T_{cz}$
limit of unity.  This may be understood as a result of the
increased transverse wandering of lines as $L_z$ increases, thus
decreasing the probability of neighboring pair reconnections.
Considering the value of $n(1)$, we see that it decreases by a
factor $\sim 2$ as $L_z$ increases from $100$ to $200$,
consistent with a random walk like behavior for the vortex line transverse
fluctuations.  This leads one to expect that in the limit $L_z\to\infty$,
for fixed $L_\perp$, the system will remain completely entangled
at all $T<T_{cz}$.

In Fig.\ \ref{fig11} we plot $n(m)$ at $T=2.4$ for systems of fixed
$L_z=100$, but varying $L_\perp=15$, $20$ and $25$.  These are systems
with a total of $N_v=9$, $16$, and $25$ field induced vortex lines
respectively.  As the maximum winding is always $m_{max}=N_v$,
the fall off of $n(m)$ at large $m$ occurs at different
$m\sim N_v$ for the different $L_\perp$.  If we normalize the
different curves in Fig.\ \ref{fig11} by $N_v$ (recall, $\sum_m n(m)=N_v$),
we find that $R\equiv n(1)/N_v$
is approaching a constant value as $L_\perp\to\infty$ (see
also Fig.\ \ref{fig7}).  Thus the fraction of disentangled lines is
approaching a well defined value.  However we show the curves without this
normalization to point up the wide intermediate region where we continue
to find $n(m)\sim 1$ as $L_\perp$ increases.  These observations suggest
that for $T<T_{cz}$
for fixed $L_z$, as $L_\perp$ increases, the equilibrium probability
distribution
of windings $n(m)/N_v$ approaches a limiting form, but the average
value of $m$ diverges.

\subsection{Vortex Ring Excitations}

We now consider the proliferation of thermally excited closed vortex ring
excitations as  illustrated in Fig.\ \ref{fig5}.  Defining
$q(p)$ as the total number of vortex rings with perimeter
$p$, we plot
in Fig.\ \ref{fig12} the log of $q(p)$ versus $1/T$, for $p=2,...$, $40$.
Our data is for the system size $L_z=100$, $L_\perp=25$.  We show the
results obtained from cooling; comparison with data from heating shows
no significant hysteresis.
For $T<T_{cz}$ the data falls along straight lines over several
orders of magnitude, clearly indicating a thermally activated form.
These lines intersect at roughly the same temperature,
$1/T_0\simeq 0.3$, thus suggesting the low temperature form
\begin{equation}
q(p)\simeq q_0{\rm e}^{-E(p)(1/T-1/T_0)}.
\label{eq:qp}
\end{equation}
In Fig.\ \ref{fig13} we plot the
the value of $E(p)$, extracted from the data of Fig.\ \ref{fig12}, versus
$p$ and find the linear dependence,
\begin{equation}
E(p)=-1.14+\varepsilon p,\qquad\qquad \varepsilon = 3.32.
\label{eq:ering}
\end{equation}
Thus for $T<T_{cz}$,
the number of rings $q(p)$ is determined by the excitation energy to
create the ring, and this energy scales linearly with the ring perimeter.
As $T$ decreases, large rings get exponentially suppressed.

For $T>T_{cz}$, we see from Fig.\ \ref{fig12}, that $q(p)$ saturates to
a constant value, and that rings on all length scales $p$ are now present.
As discussed above in connection with the winding distribution $n(m)$,
we believe that this saturation of $q(p)$ is the result of a transition
in which the vortex lines become so heavily interconnected through cuttings,
that a connected vortex tangle percolates through the entire system.
In this heavily interconnected limit, there is in general no unambiguous
way to classify a given vortex line segment as belonging to a
particular ring of size $p$, or even as belonging to a ring versus
a field induced line.
The distribution $q(p)$ would then be dominated by the
statistics of our line tracing algorithm which makes random choices
at each line cutting, rather than by any energetics.

In Fig.\ \ref{fig14} we replot our data as $q(p)$ versus $p$
for several different $T$.  We show only data for sizes $p$
in which the finite size effects, comparing different $L_\perp=15$, $20$,
$ 25$, are small
(to determine these finite size effects, we compared the
$normalized$ ring densities $q(p)/L_zL_\perp^2$).
For low $T<T_{cz}\simeq 2.6$, we see an exponential decay
$q(p)\sim {\rm exp}(-\varepsilon^\prime p/T)$, consistent with the
discussion above.
Comparison with Eqs.(\ref{eq:qp}) and (\ref{eq:ering}) gives for the
effective ring line tension, $\varepsilon^\prime$, at low temperatures,
\begin{equation}
\varepsilon^\prime=(1-T/T_0)\varepsilon .\label{eq:e1}
\end{equation}
For larger $T>T_{cz}$,
we see a slower than exponential decay, which is well fit by an
algebraic power law, $q(p)\sim p^{-x}$, $x\simeq 2.56$, $2.69$, $3.07$, $3.75$
for $T=5.0$, $3.5$, $3.0$, $2.8$ respectively.  The cross over
from exponential to algebraic decay occurs near $T_{cz}$.
The transition at $T_{cz}$ can therefore be described as the vanishing
of the ring line tension $\varepsilon^\prime$ as $T$ increases to $T_{cz}$.
This picture has some similarities with proposed vortex ring
unbinding theories of the phase transition in the ordinary three
dimensional XY model\cite{r5}.

\subsection{$2d$ Boson Analogy}

As a final indication that $T_{cz}$ is a vortex percolation-like transition,
we compute a quantity motivated by Nelson's analogy\cite{r6} between the field
induced vortex lines of a superconductor,
and the imaginary time world lines of two dimensional
bosons.  According to this analogy, the $2d$ boson superfluid density
$\rho_s$ is non-zero only when superconducting coherence parallel
to the applied magnetic field is lost\cite{r14,r8,r18}.
A convenient expression
for $\rho_s$ has been given by Cerpeley and Pollack\cite{r7} in terms of
the ``winding number" ${\bf W}$ of boson world lines,
$\rho_s=mT_{boson}\langle W^2\rangle /2\hbar^2$ where $T_{boson}$ is the
temperature of the boson system.  The mapping\cite{r6} to the superconductor
problem is given by: $\hbar/T_{boson}\to L_z$, $\hbar\to T_{super}$, $m\to
\epsilon_1\sim\pi J_0$ the single vortex line tension.
Hence $\rho_s\sim \langle W^2\rangle/L_z$.  The winding
number is defined\cite{r7} in terms of the boson world lines, or equivalently
in terms of the magnetic field induced vortex lines as,
\begin{equation}
{\bf W}={1\over L_\perp}\sum_{i=1}^{N_v}
[{\bf r}_{\perp i}(L_z)-{\bf r}_{\perp i}(0)].
\label{eq:w}
\end{equation}
{\bf W} measures the net ``winding" of the lines about the system in the
$xy$ plane ({\bf W} should not be confused
with our earlier distribution $n(m)$ which measures winding of lines
about the ${\bf\hat z}$ direction).
Since the periodic boundary condition along ${\bf\hat z}$ implies that
the set of points $\{ {\bf r}_{\perp i}(0)\}$ is equivalent to the
set of points $\{ {\bf r}_{\perp i}(L_z)\}$, ${\bf W}$
can be non-zero only if periodic boundary conditions also exist
in the ${\bf\hat x}$ and ${\bf\hat y}$ directions.
If we assume that the only vortex lines present in the system are the
magnetic field induced lines, then ${\bf W}$ is just equal to the net
vorticity in the directions perpendicular to the magnetic field,
or equivalently the perpendicular part of the $q=0$ Fourier transform of the
vortex density ${\bf n}({\bf r}_\perp,z)\equiv (1/2\pi){\bf\nabla}\times
{\bf\nabla}\theta$,
\begin{equation}
{\bf W}={1\over L_\perp}{\bf n}_{q=0}^\perp\equiv{1\over L_\perp}
[{\bf n}_{q=0}-{\bf\hat z}({\bf\hat z}\cdot{\bf n}_{q=0})].
\label{eq:w2}
\end{equation}
Note a crucial difference between ${\bf n}_{q=0}$ and the line
densities $\Delta l_\mu$ we defined earlier: $\Delta l_\mu$ measures
the total length of vortex line segments, independent of the direction
of the vorticity; ${\bf n}_{q=0}$ measures net vorticity, ie.
two line segments oriented in opposite directions will cancel in their
contribution to ${\bf n}_{q=0}$.

The Hamiltonian of our system Eq.(\ref{eq:h}) can be
expressed\cite{r2} in terms of the vortex density as,
\begin{equation}
{\cal H}[{\bf n}_q] ={2\pi^2J_0\over L_zL_\perp^2}\sum_q
({\bf n}_q-f{\bf\hat z}\delta_{q,0})\cdot ({\bf n}_q-f{\bf\hat z}\delta_{q,0})
G_q
\label{eq:hn}
\end{equation}
where the interaction $G_q\sim 1/q^2$ as $q\to 0$.
To keep the total energy finite, we
are thus rigorously constrained in our model to configurations where
${\bf n}_{q=0}^\perp=0$.
Hence as long as we assume that the only vortex lines present in the system
are the field induced lines, we must have ${\bf W}=0$.
If we now include the possibility of closed vortex ring excitations, the
identification of Eq.(\ref{eq:w2}) continues to be correct provided the rings
remain of $finite$ length $p$; the net vorticity of a finite ring always
vanishes as the vorticity must always reverse direction in order
for the ring to close back on itself.
Only if we have rings so large
(ie. infinite as $L_\perp\to\infty$)
that they wind completely around the system in the ${\bf\hat x}$ or
${\bf\hat y}$ direction, making use of the periodic boundary conditions
to close back on themselves without ever reversing the direction of
their vorticity, will the identification between ${\bf W}$
and ${\bf n}_{q=0}^\perp$ in Eq.(\ref{eq:w2})
break down.  A non-zero ${\bf W}$ computed as in Eq.(\ref{eq:w})
is now possible, provided its contribution to ${\bf n}_{q=0}^\perp$
is exactly canceled by an oppositely oriented contribution to
${\bf n}_{q=0}^\perp$ from the infinite transverse ring.

In Fig.\ \ref{fig15} we plot $\langle W^2\rangle /L_z$ versus $T$,
for system sizes
$L_z=100$, $L_\perp=15$, $20$, $25$, and $L_z=50$, $L_\perp =25$.
We see that $W^2$ is only
non-zero above $T_{cz}\simeq 2.6$.  Thus only above $T_{cz}$ do we
find vortex rings that travel completely around the system in the
direction transverse to the applied magnetic field.  This is only
possible once the vortex tangle, of interconnected magnetic field
induced lines and thermally excited rings,
percolates throughout the entire system.
Note that our results for $\langle W^2\rangle /L_z$
show some difficulties with the interpretation of this quantity
as a $2d$ boson superfluid density.
Comparing sizes $L_\perp=25$, $L_z=50$, $100$, we see no change in
$\langle W^2\rangle /L_z$, even though different $L_z$ correspond
to different temperatures $T_{boson}/\hbar  =1/L_z$ in the $2d$
boson problem.  For fixed $L_z=100$, and increasing $L_\perp$, we
see a steady decrease in $\langle W^2\rangle /L_z$ towards
zero, in contrast to expectations that the $2d$ boson $\rho_s$
should approach a finite constant.  We do not fully
understand these size dependencies.  It has been suggested\cite{r8} that the
results of Ceperley and Pollack for $\rho_s$ may not
apply in the limit of a long
range gauge interaction between $2d$ bosons, such as is the case
in our superconductor problem. Nevertheless our results continue
to support the view that $T_{cz}$ is a vortex percolation transition.

\section{Discussion}

Although we have not tried in this work to model a particular high
$T_c$ copper-oxide superconductor,
it is worth indicating in what cases our results
may qualitatively describe behavior in these materials.  Our
approximation of a uniform magnetic induction inside the material
(formally equivalent to $\lambda\to\infty$) means our
model should apply only in the limit where both the applied magnetic
field is large enough\cite{r2}
that $\lambda\gg a_v\sim\sqrt{\Phi_0/B}$, and
the Josephson coupling between
the CuO planes dominates\cite{r9} over the magnetic coupling,
$\lambda_c<\lambda^2_{ab}/d$ (where $\lambda_c$, $\lambda_{ab}$ are the
magnetic penetration lengths perpendicular to and within the CuO planes
respectively, and $d$ is the separation between the CuO planes).
Behavior in the high $T_c$ materials has been further characterized with
reference to a critical magnetic field\cite{r9,r10,r11},
$B_{cr}\simeq\Phi_0\lambda_{ab}^2/\lambda_c^2d^2$.
For $B<B_{cr}$, vortices within different CuO planes may be thought of as
correlated strings, an anisotropic Landau-Ginzburg description is
adequate, and the melting of the vortex line lattice is ``three
dimensional".  For $B>B_{cr}$, vortices within different CuO planes are
weakly coupled, the layered Lawrence-Doniach
model\cite{r12} is more appropriate,
and melting is ``quasi-two dimensional."  Since in our simulation we
have taken $J_0\sim 1/\lambda^2$ constant in all directions, and
the spacing between vortices is $a_v/a = 5 $, if we identify the
lattice constant of our numerical mesh $a$ with the spacing between
CuO planes $d$,
we have $\Phi_0/a_v^2=B<B_{cr}$ and our results apply in the region
where melting is three dimensional\cite{r13}.

One of the primary results of this paper has been to substantiate
the existence of two distinct phase transitions in our model.  At
the lower transition $T_{c\perp}$, the vortex line lattice melts upon
heating (see (I)), $\Upsilon_\perp\to 0$, and superconducting coherence
is lost in the planes perpendicular to the applied magnetic field.
At the upper transition $T_{cz}$, a vortex tangle percolates completely
through the system in the directions transverse to the applied magnetic
field, $\Upsilon_z\to 0$, and superconducting coherence is lost in
the direction parallel to the magnetic field.  Identifying the loss of
superconducting coherence with the onset of linear electric resistivity,
we therefore would expect the following experimental consequences:
As $T$ is decreased, the linear resistivity for currents applied parallel
to the magnetic field will vanish below $T_{cz}$.
However linear resistivity for currents applied perpendicular to the
magnetic field will continue to remain finite below $T_{cz}$ until
a lower $T_{c\perp}$ is reached.  This result is in agreement with
predictions by Feigel'man and co-workers\cite{r14,r8},
as well as by Glazman and Koshelev\cite{r9}, for $B<B_{cr}$.

Recent experiments\cite{r15} by Steel, White and Graybeal
on synthetic MoGe/Ge multilayers appear to show precisely
such behavior.
In these experiments, in which the magnetic field is applied perpendicular
to the layers,
the authors observe a well defined temperature
``$T_D$" at which the resistivity parallel to the magnetic field shows
a dramatic drop, accompanied by the onset of substantial nonlinearities
in the $I-V$ characteristics.  This suggests a transition where the
linear resistivity in this direction vanishes.  The resistivity perpendicular
to the magnetic field shows a kink at $T_D$, however continues to
remain linear for temperatures $T<T_D$.
Such behavior is consistent
with that of our middle phase $T_{c\perp}<T<T_{cz}$, if we identify
the experimental $T_D$ with our $T_{cz}$.  These experiments however
appear to be in the region $B>B_{cr}$, so the direct application of our
results remains unclear.

A second important result of our paper has been the observation that
upon cooling below $T_{cz}$, lines can get trapped in a disordered
entangled state where vortex line cutting is frozen out except on
long time scales.
Here the disorder is purely topological in nature\cite{r4}
and not due to any random impurities.
The importance of such entanglement on transport properties determined
by vortex line diffusion, has been stressed by Nelson
and co-workers\cite{r6,r16},
particularly with regard to pinning by large scale impurities.

Finally, we have identified the upper transition of our model, $T_{cz}$,
as the temperature at which an interconnected tangle of wandering vortex
lines and thermally excited vortex rings percolates through the system.
Our analysis of the vortex ring distribution suggestes that this is the
temperature at which the effective vortex line tension vanishes.  Since
a vortex line may qualitatively be viewed as a one dimensional
``interface" between different ground states of the $3d$ XY model, the
two transitions of our model might be viewed in analogy to the behavior
of interfaces in the $3d$ Ising model.  Our lower melting
transition $T_{c\perp}$
might be viewed like a ``roughening" transition.  Below $T_{c\perp}$
lines remain straight as $L_z\to\infty$, and are periodically ordered.
Above $T_{c\perp}$ line wandering increases with increasing $L_z$, and
lines are disordered in the plane; however lines retain a finite line tension
and so remain well defined fluctuating objects.
Our upper transition $T_{cz}$ might be
viewed as the ``bulk" transition where the effective line tension vanishes,
and detached ``bubbles," ie. vortex rings, proliferate.
A similar picture is implied in work by Bulaevskii $et$ $al$.\cite{r17},
and by Glazman and Koshelev\cite{r9}.

In our discussion of the $2d$ boson analogy in section IIIE, we
derived the important consequence that $\rho_s\equiv 0$ for all $T$
below the vortex percolation transition, from the observation that
the vortex line interaction of our model\cite{r2} was $G_q\sim 1/q^2$,
and hence energy conservation strictly requires
${\bf W}\sim {\bf n}_{q=0}^\perp =0$.  This is a direct consequence of our
approximation $\lambda\to\infty$.  For finite $\lambda$,
the interaction\cite{r2}
is $G_q\sim 1/(q^2 + \lambda^{-2})$, and now fluctuations with finite
${\bf n}_{q=0}^\perp>0$ are energetically possible in the vortex line
liquid phase.
This could imply finite $\rho_s$, and vanishing superconducting coherence
along the direction of the magnetic field.  Recent work by Tao and
Teitel\cite{r18} however shows that, for finite $\lambda$,
superconducting coherence along the magnetic field
should still persist in a hexatic vortex line liquid
state\cite{r19}, which might
exist intermediate to the vortex line lattice, and normal vortex line
liquid states.  In this case, it remains to be seen if the hexatic
to normal line liquid transition coincides with the vortex percolation
transition of our model, or if the vortex percolation remains a sharp
thermodynamic transition at all.

\section*{Acknowledgments}

It is a pleasure to thank Profs. D. R. Nelson, M. V. Feigel'man,
D. S. Fisher, and J. M. Graybeal for helpful conversations.
This work has been supported by U. S. Department of Energy grant
DE-FG02-89ER14017.

\begin{figure}
\caption{Ground state vortex line lattice for a magnetic induction
of $f=B\xi_0^2/\Phi_0=1/25$ flux quantum per unit cell of the
numerical mesh.  The view is along the direction of ${\bf B}$ and
($+$) locates the positions of the straight vortex lines.}
\label{fig1}
\end{figure}

\begin{figure}
\caption{Helicity modulus $\Upsilon_z$ along the direction of ${\bf B}$,
and $\Upsilon_\perp$ perpendicular to ${\bf B}$, for lattice sizes
$L_\perp=25$ and varying $L_z=50$, $100$, and $200$.  Both heating
and cooling are shown.  The vanishing of $\Upsilon_{z,\perp}$ indicates
two separate transitions.  No significant finite size effects are seen.}
\label{fig2}
\end{figure}

\begin{figure}
\caption{Helicity modulus $\Upsilon_z$ along the direction of ${\bf B}$,
and $\Upsilon_\perp$ perpendicular to ${\bf B}$, for lattice sizes
$L_z=25$, $L_\perp=25$ and $L_z=24$, $L_\perp=50$.  Both heating
and cooling are shown.  No significant finite size effects are seen.}
\label{fig3}
\end{figure}

\begin{figure}
\caption{Line length densities $\Delta l_\mu$, measuring absolute value
of total vortex line lengths in direction ${\bf\hat \mu}$, normalized
by total length in ground state $N_vL_z$.  For $\Delta l_z$, the ground
state line length has been subtracted, in order to show only excess
length due to fluctuations.  Both heating and cooling are shown for
a fixed lattice size $L_\perp=25$, $L_z=200$.  The solid horizontal
line indicates the total normalized length in the ground state; the
solid vertical lines mark the transition temperatures as obtained from
the vanishing of $\Upsilon_{z,\perp}$.}
\label{fig4}
\end{figure}

\begin{figure}
\caption{Schematic of possible vortex fluctuations. ($a$) shows
a directed fluctuation of a field induced vortex line; the line pierces
each plane of constant $z$ only once.  ($b$) shows a field induced
vortex line with an overhang.  ($c$) shows a closed vortex ring
excitation.}
\label{fig5}
\end{figure}

\begin{figure}
\caption{Schematic of possible reconnections of field induced
vortex lines, under application of the periodic boundary condition
in the ${\bf\hat z}$ direction.  This example shows two lines of winding
$m=1$, two groups of lines with winding $m=2$, and one group of lines
with winding $m=3$.  Solid, dashed, and dotted lines are use to distinguish
the different lines within a particular winding group.}
\label{fig6}
\end{figure}

\begin{figure}
\caption{Fraction of field induced vortex lines which are unentangled,
$R\equiv n(1)/N_v$, for various system sizes.  Note the strong hysteresis
between cooling and heating.  Entanglement increases as $L_z$ increases.}
\label{fig7}
\end{figure}

\begin{figure}
\caption{Distance $\xi_c$ along ${\bf\hat z}$ between two successive cuttings
of a single field induced vortex line.  $\xi_c\sim 2-1$ for $T>T_{cz}$
indicates a heavily interconnected vortex tangle.}
\label{fig8}
\end{figure}

\begin{figure}
\caption{Distribution of windings $n(m)$ that field induced vortex
lines make in traveling around the system along the ${\bf\hat z}$
direction.  Several different temperatures are shown for the fixed
system size $L_\perp = 15$, $L_z=200$.  For $T>T_{cz}\simeq 2.6$,
we find $n(m)\equiv 1$.}
\label{fig9}
\end{figure}

\begin{figure}
\caption{Distribution of windings $n(m)$ for fixed $T=2.4<T_{cz}$, for
system sizes $L_\perp=25$ and $L_z=50$, $100$, $200$.  As
$L_z$ increases, $n(m)\to 1$, ie. entanglement increases.}
\label{fig10}
\end{figure}

\begin{figure}
\caption{Distribution of windings $n(m)$ for fixed $T=2.4<T_{cz}$, for
system sizes $L_z=100$ and $L_\perp=15$, $20$, $25$.  Distribution remains
flat, $n(m)\simeq 1$, for wide region of intermediate $m$ as $L_\perp$
increases.}
\label{fig11}
\end{figure}

\begin{figure}
\caption{Distribution of thermally excited closed vortex
rings of perimeter $p$, versus $1/T$, for fixed system size
$L_\perp=25$, $L_z=100$.  Straight solid lines for $T<T_{cz}$
show thermally activated behavior.}
\label{fig12}
\end{figure}

\begin{figure}
\caption{Energy barrier $E(p)$ for vortex rings of perimeter $p$, as
extracted from the $T<T_{cz}$ data of Fig.\ $12$.  $E(p)$
scales linearly with $p$.}
\label{fig13}
\end{figure}

\begin{figure}
\caption{Distribution of thermally excited closed vortex rings of
perimeter $p$, versus $p$.  Several different temperatures are shown
for the fixed system size $L_\perp=25$, $L_z=100$.  For $T<T_{cz}\simeq 2.6$
solid lines are the best fit to an exponential decay.  For $T>T_{cz}$
solid lines are the best fit to an algebraic decay.}
\label{fig14}
\end{figure}

\begin{figure}
\caption{Winding number $\langle W^2\rangle /L_z$ versus temperature
for system sizes $L_z=100$, $L_\perp=15$, $20$, $25$, and $L_z=50$,
$L_\perp =25$.  $W^2>0$ only for
$T>T_{cz}\simeq 2.6$.}
\label{fig15}
\end{figure}


\begin{references}
\bibitem[*]{yl}Present Address: Department of Physics and Astronomy,
University of Rochester, Rochester NY  14627.
\bibitem{r1} Y.-H. Li and S. Teitel, Phys. Rev. Lett. {\bf 66}, 3301 (1991);
	Phys. Rev. B {\bf 45}, 5718 (1992).
\bibitem{r2} Y.-H. Li and S. Teitel, Phys. Rev. B {\bf 47}, 359 (1993).
\bibitem{r3} see also R. E. Hetzel, A. Sudb\o, and D. A. Huse, Phys. Rev. Lett.
	{\bf 69}, 518 (1992); R. Cavalcanti, G. Carneiro, and A. Gartner,
	Europhys. Lett. {\bf 17}, 449 (1992); G. Carneiro, R. Cavalcanti, and
	A. Gartner, Phys. Rev. B {\bf 47}, 5263 (1993).
\bibitem{r4} S. P. Obukhov and M. Rubinstein, Phys. Rev. Lett. {\bf 65}, 1279
	(1990); {\bf 66}, 2279 (1991).
\bibitem{r5} S. R. Shenoy, Phys. Rev. B {\bf 40}, 5056 (1989);
	B. Chattopadhyay, M. C. Mahato, and S. R. Shenoy, Phys. Rev B {\bf 47},
 	15159 (1993).
\bibitem{r6} D. R. Nelson, Phys. Rev. Lett. {\bf 60}, 1973 (1988);
	J. Stat. Phys. {\bf 57}, 511 (1989); D. R. Nelson and H. S. Seung,
	Phys. Rev. B {\bf 39}, 9153 (1989); D. R. Nelson and P. LeDoussal,
	Phys. Rev. B {\bf 42}, 10113 (1990).
\bibitem{r14} M. V. Feigel'man, Physica A {\bf 168}, 319 (1990);
	M. V. Feigel'man, V. B. Geshkenbein, and V. M. Vinokur, Pis'ma
	Zh. Eksp. Teor. Fiz. {\bf 52}, 1141 (1990) [JETP Lett. {\bf 52},
	546 (1990)].
\bibitem{r8} M. V. Feigel'man, V. B. Geshkenbein, L. B. Ioffe, and A. I.
	Larkin, preprint (1993); M. V. Feigel'man, private communication.
\bibitem{r18} T. Chen and S. Teitel, preprint (1993).
\bibitem{r7} D. M. Ceperley and E. L. Pollock, Phys. Rev. B
	{\bf 36}, 8343 (1987).
\bibitem{r9} L. I. Glazman and A. E. Koshelev, Phys. Rev. B
	{\bf 43}, 2835 (1991).
\bibitem{r10} M. V. Feigel'man, V. B. Geshkenbein, and A. I. Larkin, Physica C
	{\bf 167}, 177 (1990).
\bibitem{r11} D. S. Fisher, M. P. A. Fisher, and D. A. Huse, Phys. Rev. B
	{\bf 43}, 130 (1991).
\bibitem{r12} W. E. Lawrence and S. Doniach, in {\it Proceedings of the
	International Conference on Low Temperature Physics, LT-12}, Kyoto,
        1970, edited by E. Kanda (Keigaku, Tokyo, 1971), p. 361.
\bibitem{r13} The comment to the contrary in Ref.\ 36 of (I) is in error.
\bibitem{r15} D. G. Steel, W. R. White, and J. M. Graybeal, Phys. Rev. Lett.
	{\bf 71}, 161 (1993).
\bibitem{r16} M. C. Marchetti and D. R. Nelson, Phys. Rev. B {\bf 42}, 9938
	(1990); Physica C {\bf 174}, 40 (1991).
\bibitem{r17} L. N. Bulaevskii, M. Ledvij, and V. G. Kogan, Phys. Rev. Lett.
	{\bf 68}, 3773 (1992); L. N. Bulaevskii, S. V. Meshkov, and D. Feinberg,
	Phys. Rev. B {\bf 43}, 3728 (1991).
\bibitem{r19} M. C. Marchetti and D. R. Nelson, Phys. Rev. B {\bf 41},
	1910 (1990).
\end{references}
\end{document}